\documentclass[pra,amsmath,aps,10pt,superscriptaddress,tightenlines, twocolumn]{revtex4}
\usepackage{bm}
\usepackage{stmaryrd}
\usepackage{amsmath}
\usepackage{amssymb}
\usepackage{graphicx}
\usepackage{textcomp}
\usepackage{calrsfs}
\usepackage{yfonts}
\usepackage{amsthm}
\usepackage{lipsum}
\newcommand{\p}{{\partial}}
\newcommand{\pt}{{\partial_t}}

\newcommand{\curl}{{\nabla\times}}
\newcommand{\haf}{{\frac{1}{2}}}
\newcommand{\intf}{{\int_0^{\infty}\,}}
\newcommand{\la}{{\langle}}
\newcommand{\ra}{{\rangle}}

\begin{document}
\title{The effect of rotation on the heat transfer between two nanoparticles}

\author{Vahid Ameri}
\email{vahameri@gmail.com}
\affiliation{Department of Physics, Faculty of Science, University of Hormozgan, Bandar-Abbas, Iran}
\author{Mohammad Eghbali-Arani}
\affiliation{Department of physics, University of Kashan, Kashan, Iran}
\begin{abstract}
Quantizing the electromagnetic vacuum and medium fields of two nanoparticles, we investigate the heat transfer between them. One of the particles has been considered to rotate by angular velocity $ \omega_0 $. The effect of rotation on the absorbed heat power by the rotating nanoparticle is discussed. The results for angular velocities much smaller than the relaxation frequency $ \Gamma $ of the dielectrics are in agreement with the static nanoparticles, however increasing the angular velocity $ \omega_0 $ in comparison to the relaxation frequency of the dielectrics  $ (\omega_0\geqslant \Gamma) $  generates two sidebands in the spectrum of the absorbed heat power. The well-known near-field and far-field effects are studied and it is shown that the sidebands peaks in far-field are considerable in comparison to the main peak frequency of the spectrum.  
\end{abstract}

\maketitle
Development of nanotechnology in a wide variety of physical, chemical, biological, and medical contexts, especially in the case of rotating nanoparticles (NPs), has raised great interest. Using rotating NPs for targeting cancer cells could be one of the most important applications of them \cite{toy2014shaping,cole2011cancer,zhang2014dynamic}. Trapping and rotating NPs have been studied intensely using different methods \cite{wang2011trapping,marago2013optical,kotnala2014quantification}. Besides the important biomedical applications of rotating NPs, the effect of them also considered in some other cases e.g, on the instability of dust-acoustic waves \cite{lee2011effects}.
\par Heat transfer in the nanoscale has been studied in a variety of nanofluids \cite{kumar2004model,ben2006heat,rana2012flow}, nano to macro scales \cite{chen2002ballistic,francoeur2008near}, systems of a plane surface and NPs \cite{ameri2015radiative,dedkov2008vacuum,messina2013fluctuation,tschikin2012radiative}, between two NPs \cite{domingues2005heat}, between moving bodies \cite{volokitin2008theory}, two parallel metallic surfaces \cite{chapuis2008effects,biehs2007thermal}, two nanowires \cite{basu2013near}, and heating of NPs \cite{moran2009size}. As the rotation of nanoparticles is getting important, one can ask, how  does the rotation affect the heat transfer of NPs? It has been shown that the heat transfer absorbed by rotating NP from a plane surface could be changed \cite{ameri2015radiative}.
\par
\begin{figure}
   \includegraphics[scale=0.6]{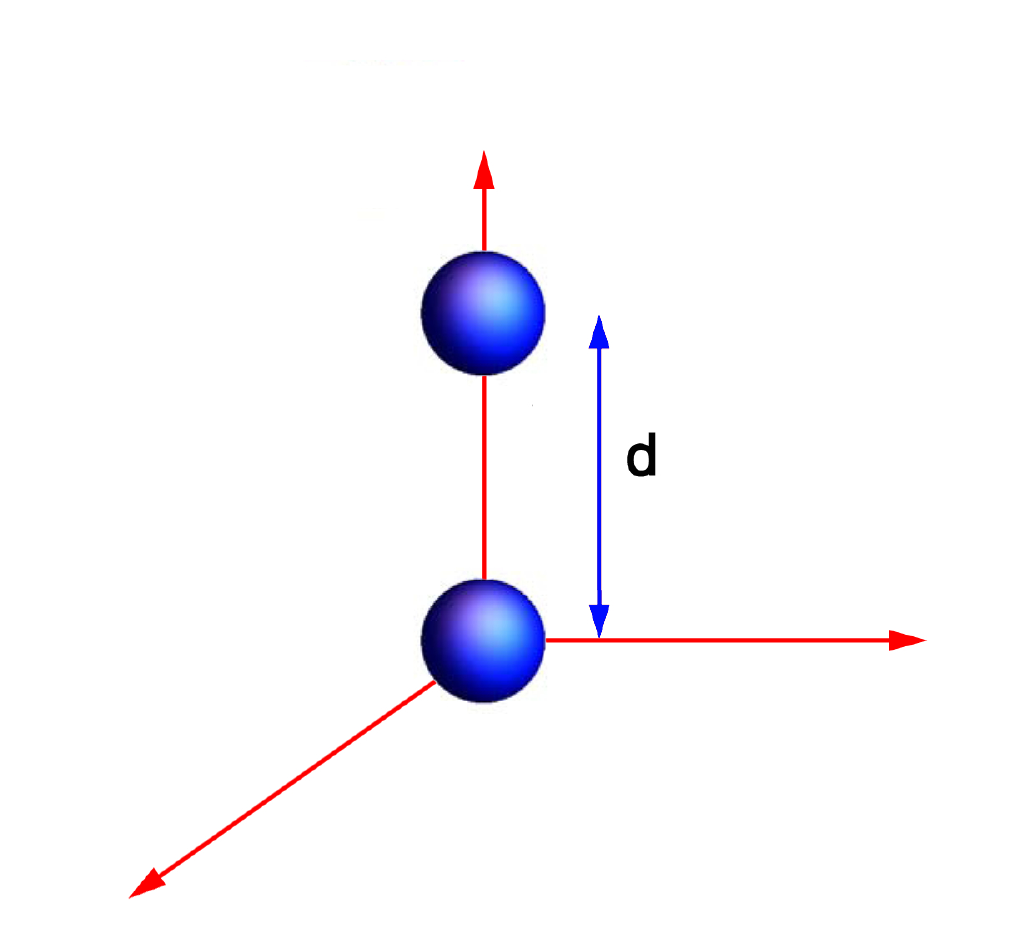}\\
  \caption{A rotating nanoparticle located on axis $ z $ a distance $ d $ from a same static NP at the origin.}\label{setup}
\end{figure}

 The aim of this work is to find the effect of rotation on the heat transfer between two NPs. To this aim, we consider a system of two NPs where one is located at the origin and the other one rotating along its axis of symmetry (axis $ z $) and located on axis $ z $ a distance $ d $ from the origin (Fig.\ref{setup}).
\par Using the canonical field quantization approach, we find the explicit form of the quantized electromagnetic and dielectric fields in the non-relativistic regime, then the heat power absorbed by rotating NP is easy to derive in this scheme. A general formula for the heat power absorbed by a rotating NP from a static NP is obtained and the effect of rotation is discussed.
\section{Electromagnetic Field Quantization}
\par The Lagrangian describing the whole system contain a term represent the electromagnetic vacuum field plus terms modelling the dielectrics and their interaction with the electromagnetic vacuum field. Following the method introduced in \cite{kheirandish2014electromagnetic,ameri2015radiative}, we study the heat transfer to the rotating NP and its physical consequences.
\par We consider the following Lagrangian for the mentioned system,
  
\begin{eqnarray}\label{L}
\mathcal{L} &=& \haf\epsilon_0\,(\pt \mathbf{A})^2-\frac{1}{2\mu_0}(\curl\mathbf{A})^2\nonumber\\
&+&\haf\intf d\nu \,[(\pt \mathbf{X}^1+\omega_0\p_{\varphi}\mathbf{X}^1)^2-\nu^2(\mathbf{X}^1)^2]\nonumber\\
&-& \epsilon_0\intf d\nu\,f_{ij}(\nu,t)X^1_j\pt A_i\nonumber\\
&+&\epsilon_0\intf d\nu\,f_{ij}(\nu,t)X^1_j (\mathbf{v}\times\curl\mathbf{A})_i\nonumber\\
&+&\haf\intf d\nu \,[(\pt \mathbf{X}^2)^2-\nu^2(\mathbf{X}^2)^2]\nonumber\\
&-& \epsilon_0\intf d\nu\,f_{ij}(\nu,0)X^2_j\pt A_i.
\end{eqnarray}
where $ X^1 $ and $ X^2 $ are the dielectric fields describing the first (rotating) and second NPs respectively and $ \omega_0 $ is the angular velocity of the rotating NP. NPs are considered to be in local thermodynamical equilibrium at temperatures $ T_1 $ and $ T_2 $ respectively. $ f_{ij}(\nu,t) $ is the coupling tensor between  the electromagnetic vacuum field and the medium fields $ X^1 $ and $ X^2 $. 
As the NPs are considered to be totally similar, they should have identical coupling tensor while to take care of the rotation of NP.1, its coupling tensor $ f_{ij}(\nu,t) $ is considered to be time dependent and for NP.2 it is time-independent as $ f_{ij}(\nu,t=0) $ where
\begin{eqnarray}\label{C}
f_{ij} (\nu,t)=\left(
  \begin{array}{ccc}
    f_{xx} (\nu) \cos(\omega_0 t) & f_{xx} (\nu) \sin(\omega_0 t) & 0 \\
    -f_{yy} (\nu) \sin(\omega_0 t) & f_{yy} (\nu) \cos(\omega_0 t) & 0 \\
    0 & 0 & f_{zz}(\nu)\\
  \end{array}
\right). \nonumber
\end{eqnarray}
To derive the Lagrangian Eq.(\ref{L}), the coordinate derivative and field transformation between rotating and fixed frames are used. As the electromagnetic fields are non-rotating fields, there is no need to modify them.
\par The response function $ \chi^{0}_{kk}(\omega) $, corresponding to set $ \omega_0=0 $, can be obtained in terms of the diagonal components of the coupling tensor $ f_{ij}(\nu,t) $ as \cite{kheirandish2014electromagnetic},
\begin{equation}\label{kapa}
\chi^{0}_{kk}(\omega) =\epsilon_0\intf d\nu\, \frac{f^{2}_{kk} (\nu)}{\nu^2-\omega^2}.
\end{equation}
The response functions of the rotating NP in the laboratory frame can be written in terms of $ \chi^{0}_{kk}(\omega) $,
\begin{eqnarray}\label{connection}
 \chi^{1}_{zz}(\omega,m)&=&\chi^{0}_{zz}(\omega-m\omega_0),\\ \nonumber
 \chi^{1}_{xx}(\omega,m)&=&\chi^{1}_{yy}(\omega,m) \\ \nonumber
 &=&\frac{1}{2}[\chi^{0}_{xx}(\omega_{+}-m\omega_0)+\chi^{0}_{xx}(\omega_{-}-m\omega_0)],\\ \nonumber
 \chi^{1}_{xy}(\omega,m)&=&-\chi^{1}_{yx}(\omega,m)
\\ \nonumber&=&\frac{1}{2i}[\chi^{0}_{xx}(\omega_{+}-m\omega_0)-\chi^{0}_{xx}(\omega_{-}-m\omega_0)],
\end{eqnarray}
where $ \omega_\pm=\omega\pm\omega_0 $.
\par Defining $P^k_{i} (\mathbf{r},t)=\epsilon_0\intf d\nu f_{ij} (\nu,t)\,X^k_j (\mathbf{r},t,\nu),$ as the electric polarization components of NPs and using Eq.(\ref{L}), we obtain the equations of motion for the electromagnetic and matter fields as
\begin{eqnarray}\label{FE}
\mathbf{P}^1 (\mathbf{r},\omega)&=&\mathbf{P}^{N,1} (\mathbf{r},\omega)+\epsilon_0 \boldsymbol{\chi}^{1} (\omega,-i\p_{\varphi})\mathbf{E}, \nonumber \\
\mathbf{P}^2 (\mathbf{r},\omega)&=&\mathbf{P}^{N,2} (\mathbf{r},\omega)+\epsilon_0 \boldsymbol{\chi}^{0} (\omega)\mathbf{E}, \nonumber \\
\Bigl\{\curl\curl &-&\frac{\omega^2}{c^2}\mathbb{I}-\frac{\omega^2}{c^2}\boldsymbol{\chi}^{1}(\omega,-i \p_\varphi) -\frac{\omega^2}{c^2}\boldsymbol{\chi}^{0}(\omega) \Bigl\}\cdot\mathbf{E}\nonumber \\
&& =\mu_0\omega^2(\mathbf{P}^{N,1}+\mathbf{P}^{N,2}),
\end{eqnarray}
where $ \mathbf{P}^{N,1} $ and $ \mathbf{P}^{N,2} $ are the fluctuating or noise electric polarizations correspond to the fluctuating or noise matter fields $ \mathbf{X}^{N,1} $ and $ \mathbf{X}^{N,2} $ for the rotating and static NPs respectively. One can expand them in terms of ladder operators as
\begin{eqnarray}\label{NField}
X^{N,1}_i (\rho,\varphi,z,\nu,t)&=&\sum_{m}\bigl[e^{i m\varphi} e^{i (\nu-m\omega_0)t}\,a^{\dag}_{i,m} (\rho,z,\nu) \nonumber\\ &+&e^{-i m\varphi}e^{-i (\nu-m\omega_0)t}\,a_{i,m} (\rho,z,\nu)\bigr],  \nonumber\\
X^{N,2}_i(\mathbf{r},\nu,t)&=&e^{i\nu t}\,b_i^{\dag}(\mathbf{r},\nu)+e^{-i\nu t}\,b_i(\mathbf{r},\nu).
\end{eqnarray}
In case of holding the NPs in thermal equilibrium at temperature $ T_1 $ and $ T_2 $, we have
\begin{eqnarray}\label{FLUC}
&&\la a^{\dag}_{i,m} (\rho,z,\nu)\,a_{j,m'} (\rho',z',\nu')\ra_T=\frac{\hbar}{4\pi\nu}\,n_{T_1} (\nu)\,\delta_{mm'}\delta_{ij}\,\nonumber \\
&&\delta(\nu-\nu')\frac{\delta(\rho-\rho')\delta(z-z')}{\rho}, \\ \nonumber
&&\la b^{\dag}_{i} (\mathbf{r},\nu),b_{j} (\mathbf{r'},\nu')\ra_T=\frac{\hbar}{2\nu}\,n_{T_2}(\nu)\, \delta_{ij}\,\delta(\nu-\nu')\delta(\mathbf{r}-\mathbf{r'}),
\end{eqnarray}
where $ n_T(\omega)=[\exp(\hbar \omega/kT)-1]^{-1} $.

\par Using (\ref{FE}) and the dyadic Green's tensor $ G_{ij} $, we find
\begin{eqnarray}\label{e}
E_i (\mathbf{r},\omega) = E_{i}^0 (\mathbf{r},\omega)+\mu_0\omega^2\int_{V_1} d\mathbf{r}'\,G_{ij} (\mathbf{r},\mathbf{r}',\omega) P^{N,1}_{j} (\mathbf{r}',\omega)\,\nonumber \\ +\mu_0\omega^2\int_{V_2} d\mathbf{r}'\,G_{ij} (\mathbf{r},\mathbf{r}',\omega) P^{N,2}_{j} (\mathbf{r}',\omega) ,
\end{eqnarray}
where the first term on the right-hand side of (\ref{e}) corresponds to the fluctuations of the electric field in electromagnetic vacuum, while the second and third terms are the induced electric field due to the fluctuations of the electric polarization of the NPs. Mitsui and Aoki reported an Observation of spontaneous quantum fluctuations in photon absorption by atoms where it is a kind of heat transfer \cite{mitsui2013observation}.
\section{Heat Transfer}
\par The rate of work done by the electromagnetic field on a differential volume $ d\mathbf{r} $ of a dielectric is given by $ \mathbf{j}\cdot (\mathbf{E}+\mathbf{v}\times \mathbf{B}) d\mathbf{r} $ where $ \mathbf{j}=\p_t\mathbf{p} - \nabla \times (\mathbf{v}\times \mathbf{p}) $ is the current density in matter. In non-relativistic regime, we ignore the terms containing velocity $ \mathbf{v} $; therefore, the radiated power of the rotating NP can be written as 
\begin{equation}\label{Power}
 \la\mathcal{P}\ra=\int_{V} d\mathbf{r}\int\int_{-\infty}^{\infty}\frac{d\omega}{2\pi}\frac{d\omega'}{2\pi}e^{-i(\omega+\omega')t}
  (i\omega)\la \mathbf{P}(\mathbf{r},\omega)\cdot\mathbf{E}(\mathbf{r},\omega')\ra.
\end{equation}
The radiated power $ \la\mathcal{P}\ra $ contains all emitted and absorbed energy of  the rotating NP due to the interaction with electric field $ \mathbf{E} $. One of the terms contains the fluctuating electrical polarization of the static NP at the origin $ \la \mathbf{p}^{N,2}_i(\mathbf{r},\omega)\cdot \mathbf{p}^{N,2}_j(\mathbf{r}',\omega')\ra$, which we will focus on in the following, is responsible for the heat transfer power $ \la\mathcal{P}\ra_{HT} $ from the static NP at the origin to the rotating NP on the axis $ z $ a distance $ d $ from the origin. Using Eqs. (\ref{e}), and (\ref{Power}), we derive
\begin{eqnarray}\label{pht}
\la\mathcal{P}\ra_{HT}=\frac{2\hbar}{\pi}\intf d\omega \frac{\omega^5}{c^4}\mbox{Im}\,[\alpha_{ij}^1(\omega)]\mbox{Im}\,[\alpha_{kk}^0(\omega)]n_{T_2}(\omega) \nonumber \\
\times \int d\mathbf{r}' G_{ik}(\mathbf{r},\mathbf{r}',\omega)G_{kj}^*(\mathbf{r},\mathbf{r}',\omega).
\end{eqnarray}
where $ \mbox{Im}\,[\alpha_{ij}(\omega)]=V\mbox{Im}\,[\chi_{ij}(\omega,m=0)] $ and $ V $ represents the volume of NPs.
One can find a proper dyadic Green's tensor $ G_{ij} $ for Eq.(\ref{e}) as
\begin{eqnarray}\label{g}
G_{ij}(\mathbf{r},\mathbf{r}',\omega)=\frac{e^{ikR}}{R^3k^2}[(k^2R^2+ikR-1)\delta_{ij}\nonumber \\
-(k^2R^2+3ikR-3)\frac{R_iR_j}{R^2}],
\end{eqnarray}
where $ \mathbf{R}=\mathbf{r}-\mathbf{r}' $ and $ k=\omega/c $.
In this work, the interparticle distance $ d $ between NPs are chosen to be much bigger than their radius, therefore, it is a good approximation to consider them as point-like particles. On the other hand, using this approximation, the components of dyadic Green's tensor $ G_{ij} $ can be simplified to
\begin{eqnarray}
G_{xx}(0,d\hat{z},\omega)&=&G_{yy}(0,d\hat{z},\omega)=\frac{e^{ikd}}{d^3k^2}(k^2d^2+ikd-1), \nonumber \\
G_{zz}(0,d\hat{z},\omega)&=&\frac{2e^{ikd}}{d^3k^2}(1-ikd),
\end{eqnarray}  
and all other components are vanished.

\section{Discussion and Results}
\par To find some numerical results, the NPs are considered to be made of Silicon Carbide (SiC) with similar radius $ (a=2nm) $, where the dielectric function is given by the oscillator model \cite{palik1998handbook},
\begin{equation}
\varepsilon(\omega)=\varepsilon_\infty(1+\frac{\omega_L^2-\omega_T^2 }{\omega_T^2-\omega^2-i\Gamma \omega }),
\end{equation}
with $\varepsilon_\infty=6.7$, $\omega_L=1.823\times 10^{14}\mbox{rad/s}$, $\omega_T=1.492\times 10^{14}\mbox{rad/s}$, and $\Gamma = 8.954\times 10^{11}\mbox{rad/s}$.
\begin{figure}
   \includegraphics[scale=0.5]{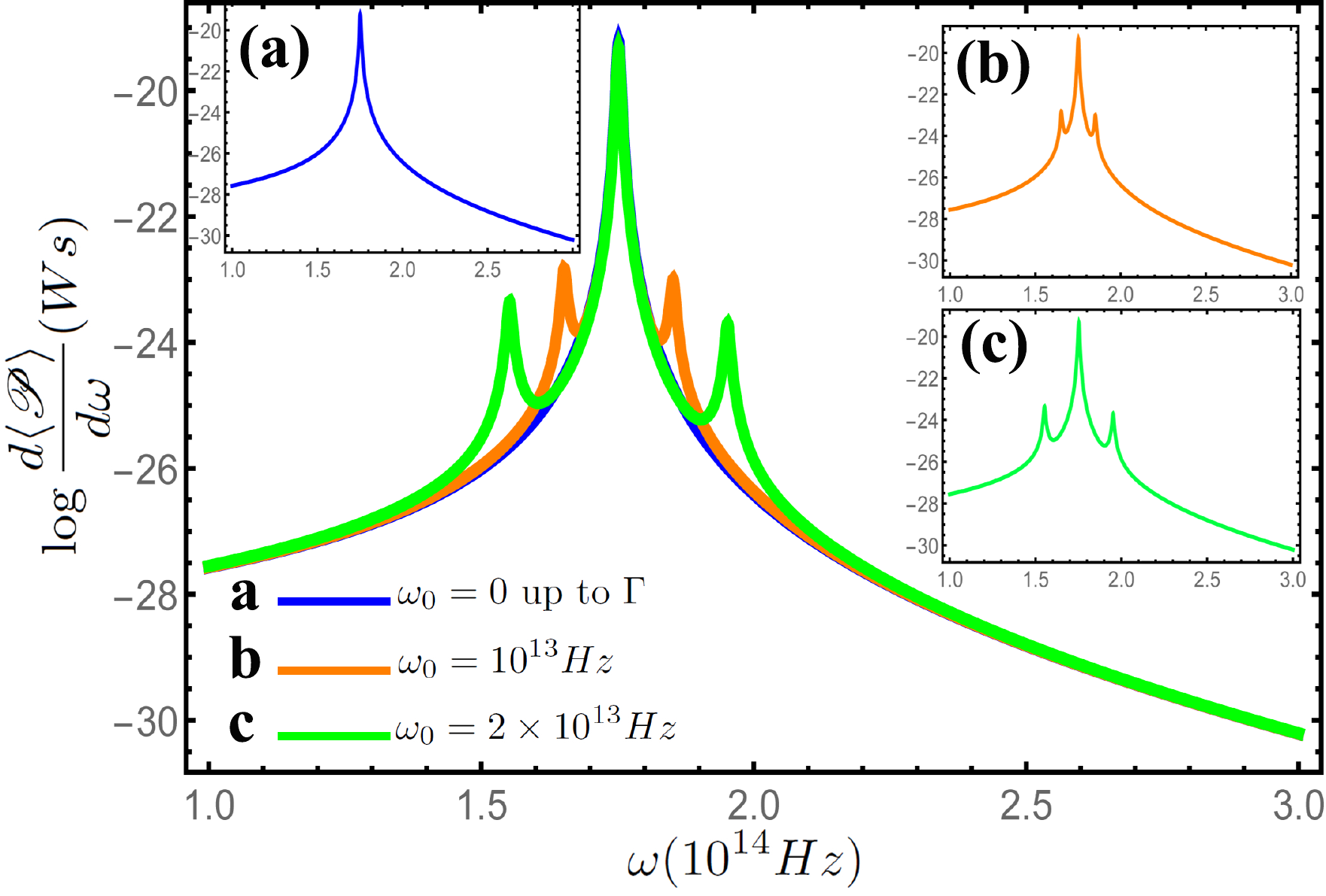}\\
  \caption{ Logarithmic (base 10) plot of the absorbed heat power spectrum of a rotating NP located on the axis $ z $ a distance $ d=10\mbox{nm} $ (near-field) from a static NP at the origin as a function of frequency where the static NP is considered to be at room temperature $ (T=300\mbox{K}) $. }\label{sw0}
\end{figure}
\begin{figure}
   \includegraphics[scale=0.75]{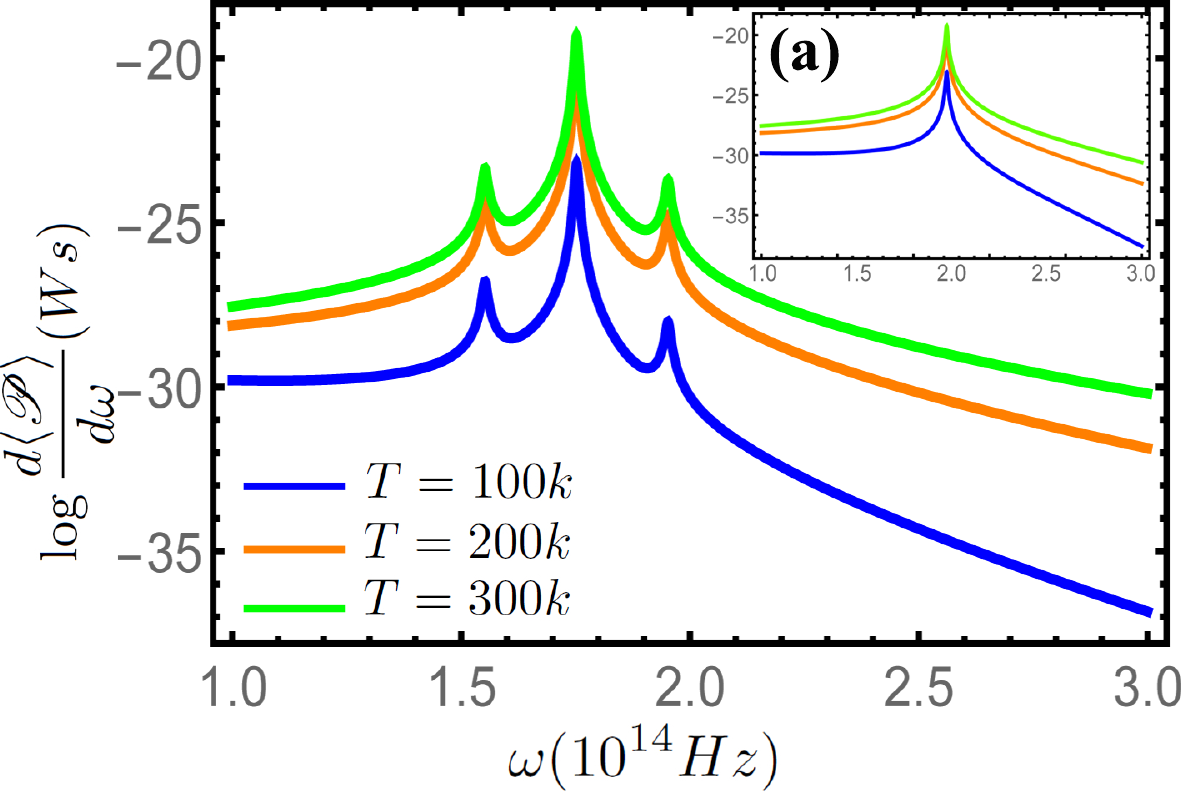}\\
  \caption{ Logarithmic (base 10) plot of the absorbed heat power spectrum of a rotating NP $ \omega_0=2\times 10^{13} \mbox{rad/s}$, (a) a non-rotating NP, located on the axis $ z $ a distance $ d=10\mbox{nm} $ from a static NP at the origin as a function of frequency for different temperature of static NP.}\label{st}
\end{figure}
\begin{figure}
   \includegraphics[scale=0.55]{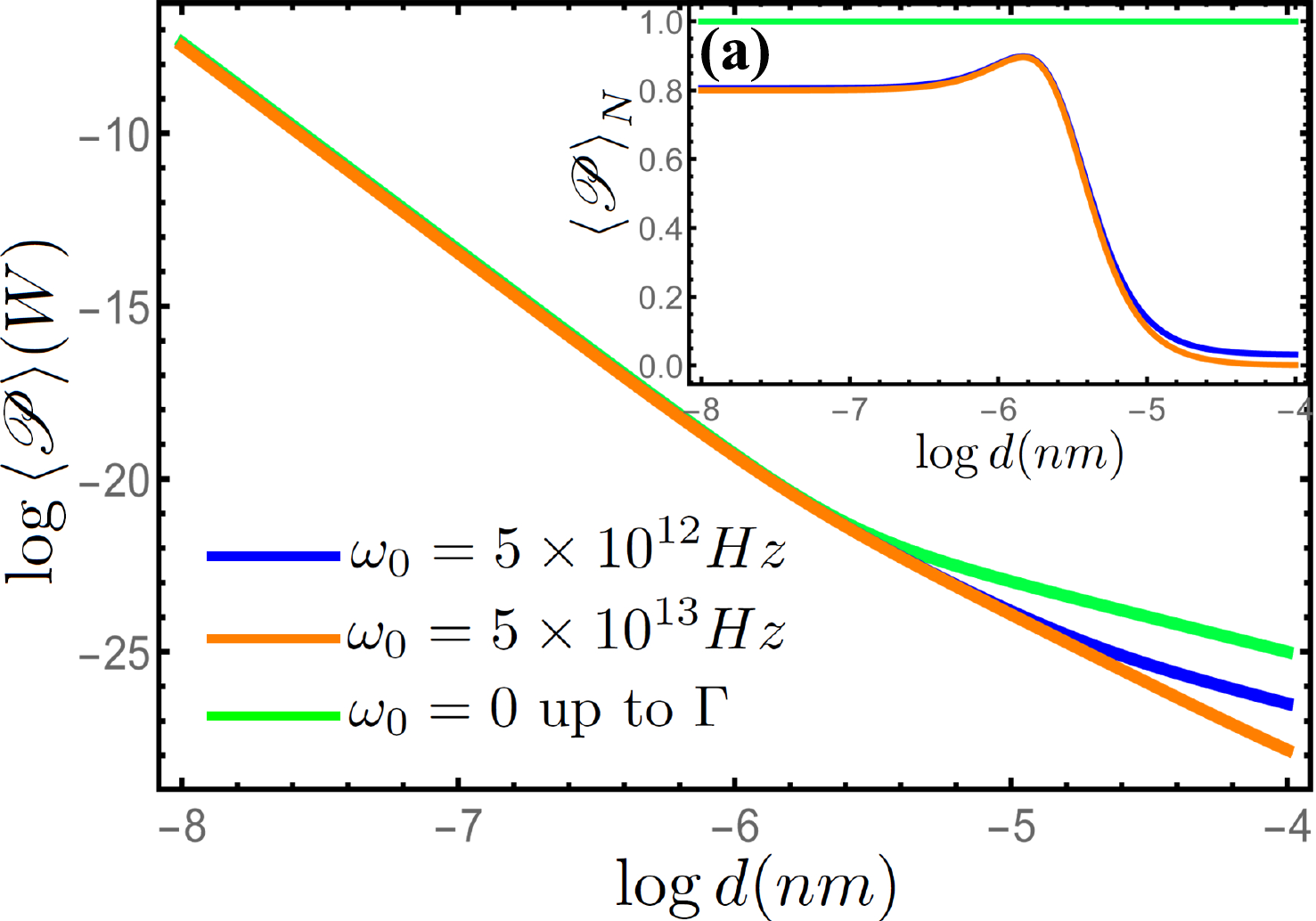}\\
  \caption{ Logarithmic (base 10) plot of total absorbed heat power by the rotating NP with angular velocity $ \omega_0 $ , (a) total absorbed heat power by the rotating NP with angular velocity $ \omega_0 $ divided by the total absorbed heat power of non-rotating NP, as a function of the interparticle distance $ d $ where the static NP is considered to be at room temperature $ (T=300\mbox{K}) $.   }\label{sz}
\end{figure}
\begin{figure}
   \includegraphics[scale=0.55]{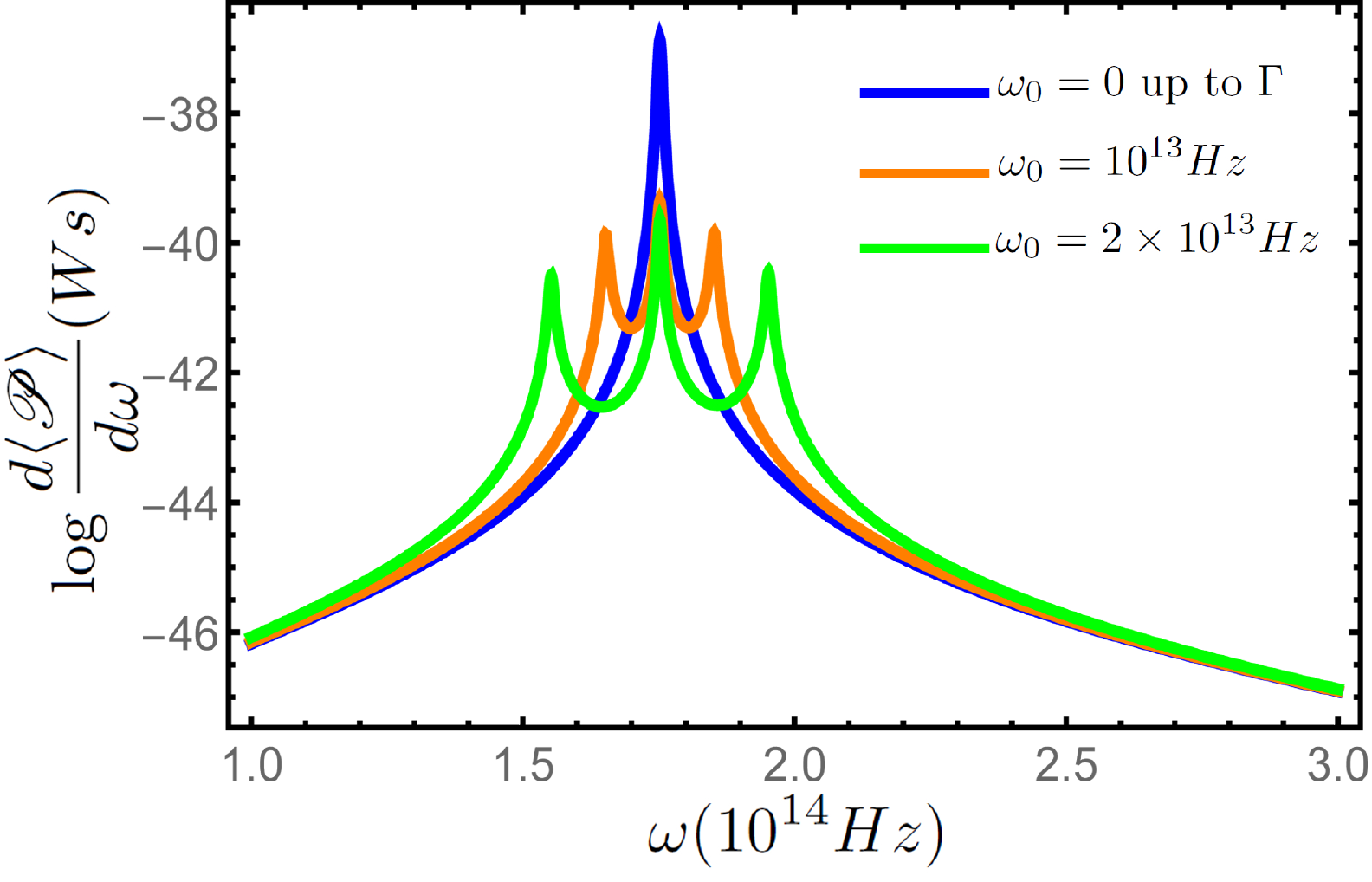}\\
  \caption{ Logarithmic (base 10) plot of the absorbed heat power spectrum of a rotating NP located on the axis $ z $ a distance $ d=10\mbox{mm} $ (far-field) from a static NP at the origin as a function of frequency where the static NP is considered to be at room temperature $ (T=300\mbox{K}) $. }\label{sw0f}
\end{figure}
\par It has been seen that in a cavity optomechanics, a mirror oscillating with frequency $ \omega_0 $, the optical sidebands are created around the incoming light frequency $ \omega $,
\begin{equation}\label{wp}
\omega'=\omega\pm\omega_0
\end{equation}
although, in this work we are far from the cavity optomechanics, but both cases contain a kind of oscillation frequency $ \omega_0 $ on the matter fields. Thus, one can expect identical results on the spectrum of the absorbed heat transfer, where interestingly, it is supported by Fig. (\ref{sw0}). It shows that, increasing the angular velocity $ \omega_0 $, can cause a couple of sidebands on the spectrum of the absorbed heat transfer by rotating NP from the static NP, where the sidebands appeared around the remarkable peak frequency $ \omega $ of the absorbed heat transfer spectrum of non-rotating NPs. The sidebands frequencies are given by Eq. (\ref{wp}). While the spectrum, for angular velocities smaller than the relaxation frequency $ \Gamma $ of dielectrics, is the same as the spectrum for non-rotating NPs. Having a discontinuous behavior for $ \omega_0<\Gamma $ and $ \omega_0>\Gamma $ is not surprising. In fact, the relaxation frequency $ \Gamma $ is well related to the response time of the dielectric to an external electric field. So any change that happens faster than this response time may cause new behaviors. 
\par The effect of static NP temperature $ T_2 $ on the spectrum of the absorbed heat transfer of rotating Np has been depicted in Fig. (\ref{st}). As a result of that, the temperature of the static NP will affect the absorbed heat transfer of rotating and non-rotating NP in the similar way.
\par To focus more on the effect of rotation in the near and far field of the static NP, the total absorbed heat power of rotating NP as a function of the interparticle distance $ d $ has been depicted in Fig. (\ref{sz}), where it provides a dependence on $ d^{-6} $ in the near field and on $ d^{-2} $ in the far field  as reported previously for non-rotating NPs \cite{domingues2005heat, manjavacas2012radiative}but interestingly, for $ \omega_0=5\times 10^{13} $, in the mean while between the near field to the far field, interparticle distances $ 10^{-6}m $ to $  10^{-4}m $, the total absorbed heat power of rotating NP depends on $ d^{-4} $. In case of non-rotating NPs, as one can see in Fig. (\ref{sz}), there is no such effect and also no one has reported such functionality of total absorbed heat power of a NP. 
\par Fig. (\ref{sz}a) shows a transition on the total absorbed heat power of rotating NP normalized by the total absorbed heat power of a non-rotating NP. Also, there is a remarkable maximum around one micron on that.As it appeared where the interparticle distance equals to the wavelength of the peak frequency of the absorbed radiation spectrum, it might be the effect of a resonance. As a consequence of transition between near field and far field of the static NP, One can expect, the effect of rotation on the absorbed heat power of a rotating NP can be quite different in near field and far field of the static Np. Therefore, we renew the plot of Fig. (\ref{sw0}), where it was depicted in the near field, in far field of the static NP. In Fig. (\ref{sw0f}) surprisingly, the contribution of sidebands peaks has raised in compare to the main peak of the spectrum. This future may find a lot of applications e.g in designing the new optomechanical systems.
\section{Conclusions}
\par In conclusion, the effect of rotation on the heat transfer between two NPs has been analysed. Two sidebands appeared around the peak frequency of the heat transfer spectrum of non-rotating NPs. While the peaks of the sidebands are so small in compare to the main peak in the near field, they are considerable in the far field even up to the same order of magnitude in compare to the main peak frequency of the spectrum. It has been shown that the total absorbed heat power of a rotating NP, with angular velocities $ \omega_o $ larger than the relaxation frequency $ \Gamma $ of the dielectric, experience a new regime between the near field and the far field where it provides the dependence on $ d^{-4} $ and the range of this interesting regime decreases by the angular velocity of the NP and finally vanishes for the angular velocities smaller than the relaxation frequency $ \Gamma $ of the dielectric.   

\section*{Acknowledgement}
The authors are grateful to University of Hormozgan and University of Kashan for the support of this work.
\section*{Author contributed statement}
 Both authors contributed equally to the paper.  
\bibliography{sci}

\begin{thebibliography}{26}%
\makeatletter
\providecommand \@ifxundefined [1]{%
 \@ifx{#1\undefined}
}%
\providecommand \@ifnum [1]{%
 \ifnum #1\expandafter \@firstoftwo
 \else \expandafter \@secondoftwo
 \fi
}%
\providecommand \@ifx [1]{%
 \ifx #1\expandafter \@firstoftwo
 \else \expandafter \@secondoftwo
 \fi
}%
\providecommand \natexlab [1]{#1}%
\providecommand \enquote  [1]{``#1''}%
\providecommand \bibnamefont  [1]{#1}%
\providecommand \bibfnamefont [1]{#1}%
\providecommand \citenamefont [1]{#1}%
\providecommand \href@noop [0]{\@secondoftwo}%
\providecommand \href [0]{\begingroup \@sanitize@url \@href}%
\providecommand \@href[1]{\@@startlink{#1}\@@href}%
\providecommand \@@href[1]{\endgroup#1\@@endlink}%
\providecommand \@sanitize@url [0]{\catcode `\\12\catcode `\$12\catcode
  `\&12\catcode `\#12\catcode `\^12\catcode `\_12\catcode `\%12\relax}%
\providecommand \@@startlink[1]{}%
\providecommand \@@endlink[0]{}%
\providecommand \url  [0]{\begingroup\@sanitize@url \@url }%
\providecommand \@url [1]{\endgroup\@href {#1}{\urlprefix }}%
\providecommand \urlprefix  [0]{URL }%
\providecommand \Eprint [0]{\href }%
\providecommand \doibase [0]{http://dx.doi.org/}%
\providecommand \selectlanguage [0]{\@gobble}%
\providecommand \bibinfo  [0]{\@secondoftwo}%
\providecommand \bibfield  [0]{\@secondoftwo}%
\providecommand \translation [1]{[#1]}%
\providecommand \BibitemOpen [0]{}%
\providecommand \bibitemStop [0]{}%
\providecommand \bibitemNoStop [0]{.\EOS\space}%
\providecommand \EOS [0]{\spacefactor3000\relax}%
\providecommand \BibitemShut  [1]{\csname bibitem#1\endcsname}%
\let\auto@bib@innerbib\@empty
\bibitem [{\citenamefont {Toy}\ \emph {et~al.}(2014)\citenamefont {Toy},
  \citenamefont {Peiris}, \citenamefont {Ghaghada},\ and\ \citenamefont
  {Karathanasis}}]{toy2014shaping}%
  \BibitemOpen
  \bibfield  {author} {\bibinfo {author} {\bibfnamefont {R.}~\bibnamefont
  {Toy}}, \bibinfo {author} {\bibfnamefont {P.~M.}\ \bibnamefont {Peiris}},
  \bibinfo {author} {\bibfnamefont {K.~B.}\ \bibnamefont {Ghaghada}}, \ and\
  \bibinfo {author} {\bibfnamefont {E.}~\bibnamefont {Karathanasis}},\
  }\href@noop {} {\bibfield  {journal} {\bibinfo  {journal} {Nanomedicine}\
  }\textbf {\bibinfo {volume} {9}},\ \bibinfo {pages} {121} (\bibinfo {year}
  {2014})}\BibitemShut {NoStop}%
\bibitem [{\citenamefont {Cole}\ \emph {et~al.}(2011)\citenamefont {Cole},
  \citenamefont {Yang},\ and\ \citenamefont {David}}]{cole2011cancer}%
  \BibitemOpen
  \bibfield  {author} {\bibinfo {author} {\bibfnamefont {A.~J.}\ \bibnamefont
  {Cole}}, \bibinfo {author} {\bibfnamefont {V.~C.}\ \bibnamefont {Yang}}, \
  and\ \bibinfo {author} {\bibfnamefont {A.~E.}\ \bibnamefont {David}},\
  }\href@noop {} {\bibfield  {journal} {\bibinfo  {journal} {Trends in
  biotechnology}\ }\textbf {\bibinfo {volume} {29}},\ \bibinfo {pages} {323}
  (\bibinfo {year} {2011})}\BibitemShut {NoStop}%
\bibitem [{\citenamefont {Zhang}\ \emph {et~al.}(2014)\citenamefont {Zhang},
  \citenamefont {Kircher}, \citenamefont {Koch}, \citenamefont {Eliasson},
  \citenamefont {Goldberg},\ and\ \citenamefont
  {Renström}}]{zhang2014dynamic}%
  \BibitemOpen
  \bibfield  {author} {\bibinfo {author} {\bibfnamefont {E.}~\bibnamefont
  {Zhang}}, \bibinfo {author} {\bibfnamefont {M.~F.}\ \bibnamefont {Kircher}},
  \bibinfo {author} {\bibfnamefont {M.}~\bibnamefont {Koch}}, \bibinfo {author}
  {\bibfnamefont {L.}~\bibnamefont {Eliasson}}, \bibinfo {author}
  {\bibfnamefont {S.~N.}\ \bibnamefont {Goldberg}}, \ and\ \bibinfo {author}
  {\bibfnamefont {E.}~\bibnamefont {Renström}},\ }\href@noop {} {\bibfield
  {journal} {\bibinfo  {journal} {ACS nano}\ }\textbf {\bibinfo {volume} {8}},\
  \bibinfo {pages} {3192} (\bibinfo {year} {2014})}\BibitemShut {NoStop}%
\bibitem [{\citenamefont {Wang}\ \emph {et~al.}(2011)\citenamefont {Wang},
  \citenamefont {Schonbrun}, \citenamefont {Steinvurzel},\ and\ \citenamefont
  {Crozier}}]{wang2011trapping}%
  \BibitemOpen
  \bibfield  {author} {\bibinfo {author} {\bibfnamefont {K.}~\bibnamefont
  {Wang}}, \bibinfo {author} {\bibfnamefont {E.}~\bibnamefont {Schonbrun}},
  \bibinfo {author} {\bibfnamefont {P.}~\bibnamefont {Steinvurzel}}, \ and\
  \bibinfo {author} {\bibfnamefont {K.~B.}\ \bibnamefont {Crozier}},\
  }\href@noop {} {\bibfield  {journal} {\bibinfo  {journal} {Nature
  communications}\ }\textbf {\bibinfo {volume} {2}},\ \bibinfo {pages} {469}
  (\bibinfo {year} {2011})}\BibitemShut {NoStop}%
\bibitem [{\citenamefont {Marag{\`o}}\ \emph {et~al.}(2013)\citenamefont
  {Marag{\`o}}, \citenamefont {Jones}, \citenamefont {Gucciardi}, \citenamefont
  {Volpe},\ and\ \citenamefont {Ferrari}}]{marago2013optical}%
  \BibitemOpen
  \bibfield  {author} {\bibinfo {author} {\bibfnamefont {O.~M.}\ \bibnamefont
  {Marag{\`o}}}, \bibinfo {author} {\bibfnamefont {P.~H.}\ \bibnamefont
  {Jones}}, \bibinfo {author} {\bibfnamefont {P.~G.}\ \bibnamefont
  {Gucciardi}}, \bibinfo {author} {\bibfnamefont {G.}~\bibnamefont {Volpe}}, \
  and\ \bibinfo {author} {\bibfnamefont {A.~C.}\ \bibnamefont {Ferrari}},\
  }\href@noop {} {\bibfield  {journal} {\bibinfo  {journal} {Nature
  nanotechnology}\ }\textbf {\bibinfo {volume} {8}},\ \bibinfo {pages} {807}
  (\bibinfo {year} {2013})}\BibitemShut {NoStop}%
\bibitem [{\citenamefont {Kotnala}\ and\ \citenamefont
  {Gordon}(2014)}]{kotnala2014quantification}%
  \BibitemOpen
  \bibfield  {author} {\bibinfo {author} {\bibfnamefont {A.}~\bibnamefont
  {Kotnala}}\ and\ \bibinfo {author} {\bibfnamefont {R.}~\bibnamefont
  {Gordon}},\ }\href@noop {} {\bibfield  {journal} {\bibinfo  {journal} {Nano
  letters}\ }\textbf {\bibinfo {volume} {14}},\ \bibinfo {pages} {853}
  (\bibinfo {year} {2014})}\BibitemShut {NoStop}%
\bibitem [{\citenamefont {Lee}(2011)}]{lee2011effects}%
  \BibitemOpen
  \bibfield  {author} {\bibinfo {author} {\bibfnamefont {M.-J.}\ \bibnamefont
  {Lee}},\ }\href@noop {} {\bibfield  {journal} {\bibinfo  {journal} {Journal
  of nanoscience and nanotechnology}\ }\textbf {\bibinfo {volume} {11}},\
  \bibinfo {pages} {880} (\bibinfo {year} {2011})}\BibitemShut {NoStop}%
\bibitem [{\citenamefont {Kumar}\ \emph {et~al.}(2004)\citenamefont {Kumar},
  \citenamefont {Patel}, \citenamefont {Kumar}, \citenamefont {Sundararajan},
  \citenamefont {Pradeep},\ and\ \citenamefont {Das}}]{kumar2004model}%
  \BibitemOpen
  \bibfield  {author} {\bibinfo {author} {\bibfnamefont {D.~H.}\ \bibnamefont
  {Kumar}}, \bibinfo {author} {\bibfnamefont {H.~E.}\ \bibnamefont {Patel}},
  \bibinfo {author} {\bibfnamefont {V.~R.}\ \bibnamefont {Kumar}}, \bibinfo
  {author} {\bibfnamefont {T.}~\bibnamefont {Sundararajan}}, \bibinfo {author}
  {\bibfnamefont {T.}~\bibnamefont {Pradeep}}, \ and\ \bibinfo {author}
  {\bibfnamefont {S.~K.}\ \bibnamefont {Das}},\ }\href@noop {} {\bibfield
  {journal} {\bibinfo  {journal} {Physical Review Letters}\ }\textbf {\bibinfo
  {volume} {93}},\ \bibinfo {pages} {144301} (\bibinfo {year}
  {2004})}\BibitemShut {NoStop}%
\bibitem [{\citenamefont {Ben-Abdallah}(2006)}]{ben2006heat}%
  \BibitemOpen
  \bibfield  {author} {\bibinfo {author} {\bibfnamefont {P.}~\bibnamefont
  {Ben-Abdallah}},\ }\href@noop {} {\bibfield  {journal} {\bibinfo  {journal}
  {Applied physics letters}\ }\textbf {\bibinfo {volume} {89}},\ \bibinfo
  {pages} {113117} (\bibinfo {year} {2006})}\BibitemShut {NoStop}%
\bibitem [{\citenamefont {Rana}\ and\ \citenamefont
  {Bhargava}(2012)}]{rana2012flow}%
  \BibitemOpen
  \bibfield  {author} {\bibinfo {author} {\bibfnamefont {P.}~\bibnamefont
  {Rana}}\ and\ \bibinfo {author} {\bibfnamefont {R.}~\bibnamefont
  {Bhargava}},\ }\href@noop {} {\bibfield  {journal} {\bibinfo  {journal}
  {Communications in Nonlinear Science and Numerical Simulation}\ }\textbf
  {\bibinfo {volume} {17}},\ \bibinfo {pages} {212} (\bibinfo {year}
  {2012})}\BibitemShut {NoStop}%
\bibitem [{\citenamefont {Chen}(2002)}]{chen2002ballistic}%
  \BibitemOpen
  \bibfield  {author} {\bibinfo {author} {\bibfnamefont {G.}~\bibnamefont
  {Chen}},\ }\href@noop {} {\bibfield  {journal} {\bibinfo  {journal} {Journal
  of Heat transfer}\ }\textbf {\bibinfo {volume} {124}},\ \bibinfo {pages}
  {320} (\bibinfo {year} {2002})}\BibitemShut {NoStop}%
\bibitem [{\citenamefont {Francoeur}\ \emph {et~al.}(2008)\citenamefont
  {Francoeur}, \citenamefont {Meng{\"u}{\c{c}}},\ and\ \citenamefont
  {Vaillon}}]{francoeur2008near}%
  \BibitemOpen
  \bibfield  {author} {\bibinfo {author} {\bibfnamefont {M.}~\bibnamefont
  {Francoeur}}, \bibinfo {author} {\bibfnamefont {M.~P.}\ \bibnamefont
  {Meng{\"u}{\c{c}}}}, \ and\ \bibinfo {author} {\bibfnamefont
  {R.}~\bibnamefont {Vaillon}},\ }\href@noop {} {\bibfield  {journal} {\bibinfo
   {journal} {Applied Physics Letters}\ }\textbf {\bibinfo {volume} {93}},\
  \bibinfo {pages} {043109} (\bibinfo {year} {2008})}\BibitemShut {NoStop}%
\bibitem [{\citenamefont {Ameri}\ \emph {et~al.}(2015)\citenamefont {Ameri},
  \citenamefont {Aporvari},\ and\ \citenamefont
  {Kheirandish}}]{ameri2015radiative}%
  \BibitemOpen
  \bibfield  {author} {\bibinfo {author} {\bibfnamefont {V.}~\bibnamefont
  {Ameri}}, \bibinfo {author} {\bibfnamefont {M.~S.}\ \bibnamefont {Aporvari}},
  \ and\ \bibinfo {author} {\bibfnamefont {F.}~\bibnamefont {Kheirandish}},\
  }\href@noop {} {\bibfield  {journal} {\bibinfo  {journal} {Physical Review
  A}\ }\textbf {\bibinfo {volume} {92}},\ \bibinfo {pages} {022110} (\bibinfo
  {year} {2015})}\BibitemShut {NoStop}%
\bibitem [{\citenamefont {Dedkov}\ and\ \citenamefont
  {Kyasov}(2008)}]{dedkov2008vacuum}%
  \BibitemOpen
  \bibfield  {author} {\bibinfo {author} {\bibfnamefont {G.}~\bibnamefont
  {Dedkov}}\ and\ \bibinfo {author} {\bibfnamefont {A.}~\bibnamefont
  {Kyasov}},\ }\href@noop {} {\bibfield  {journal} {\bibinfo  {journal}
  {Journal of Physics: Condensed Matter}\ }\textbf {\bibinfo {volume} {20}},\
  \bibinfo {pages} {354006} (\bibinfo {year} {2008})}\BibitemShut {NoStop}%
\bibitem [{\citenamefont {Messina}\ \emph {et~al.}(2013)\citenamefont
  {Messina}, \citenamefont {Tschikin}, \citenamefont {Biehs},\ and\
  \citenamefont {Ben-Abdallah}}]{messina2013fluctuation}%
  \BibitemOpen
  \bibfield  {author} {\bibinfo {author} {\bibfnamefont {R.}~\bibnamefont
  {Messina}}, \bibinfo {author} {\bibfnamefont {M.}~\bibnamefont {Tschikin}},
  \bibinfo {author} {\bibfnamefont {S.-A.}\ \bibnamefont {Biehs}}, \ and\
  \bibinfo {author} {\bibfnamefont {P.}~\bibnamefont {Ben-Abdallah}},\
  }\href@noop {} {\bibfield  {journal} {\bibinfo  {journal} {Physical Review
  B}\ }\textbf {\bibinfo {volume} {88}},\ \bibinfo {pages} {104307} (\bibinfo
  {year} {2013})}\BibitemShut {NoStop}%
\bibitem [{\citenamefont {Tschikin}\ \emph {et~al.}(2012)\citenamefont
  {Tschikin}, \citenamefont {Biehs}, \citenamefont {Rosa},\ and\ \citenamefont
  {Ben-Abdallah}}]{tschikin2012radiative}%
  \BibitemOpen
  \bibfield  {author} {\bibinfo {author} {\bibfnamefont {M.}~\bibnamefont
  {Tschikin}}, \bibinfo {author} {\bibfnamefont {S.-A.}\ \bibnamefont {Biehs}},
  \bibinfo {author} {\bibfnamefont {F.~S.}\ \bibnamefont {Rosa}}, \ and\
  \bibinfo {author} {\bibfnamefont {P.}~\bibnamefont {Ben-Abdallah}},\
  }\href@noop {} {\bibfield  {journal} {\bibinfo  {journal} {The European
  Physical Journal B-Condensed Matter and Complex Systems}\ }\textbf {\bibinfo
  {volume} {85}},\ \bibinfo {pages} {1} (\bibinfo {year} {2012})}\BibitemShut
  {NoStop}%
\bibitem [{\citenamefont {Domingues}\ \emph {et~al.}(2005)\citenamefont
  {Domingues}, \citenamefont {Volz}, \citenamefont {Joulain},\ and\
  \citenamefont {Greffet}}]{domingues2005heat}%
  \BibitemOpen
  \bibfield  {author} {\bibinfo {author} {\bibfnamefont {G.}~\bibnamefont
  {Domingues}}, \bibinfo {author} {\bibfnamefont {S.}~\bibnamefont {Volz}},
  \bibinfo {author} {\bibfnamefont {K.}~\bibnamefont {Joulain}}, \ and\
  \bibinfo {author} {\bibfnamefont {J.-J.}\ \bibnamefont {Greffet}},\
  }\href@noop {} {\bibfield  {journal} {\bibinfo  {journal} {Physical review
  letters}\ }\textbf {\bibinfo {volume} {94}},\ \bibinfo {pages} {085901}
  (\bibinfo {year} {2005})}\BibitemShut {NoStop}%
\bibitem [{\citenamefont {Volokitin}\ and\ \citenamefont
  {Persson}(2008)}]{volokitin2008theory}%
  \BibitemOpen
  \bibfield  {author} {\bibinfo {author} {\bibfnamefont {A.}~\bibnamefont
  {Volokitin}}\ and\ \bibinfo {author} {\bibfnamefont {B.}~\bibnamefont
  {Persson}},\ }\href@noop {} {\bibfield  {journal} {\bibinfo  {journal}
  {Physical Review B}\ }\textbf {\bibinfo {volume} {78}},\ \bibinfo {pages}
  {155437} (\bibinfo {year} {2008})}\BibitemShut {NoStop}%
\bibitem [{\citenamefont {Chapuis}\ \emph {et~al.}(2008)\citenamefont
  {Chapuis}, \citenamefont {Volz}, \citenamefont {Henkel}, \citenamefont
  {Joulain},\ and\ \citenamefont {Greffet}}]{chapuis2008effects}%
  \BibitemOpen
  \bibfield  {author} {\bibinfo {author} {\bibfnamefont {P.-O.}\ \bibnamefont
  {Chapuis}}, \bibinfo {author} {\bibfnamefont {S.}~\bibnamefont {Volz}},
  \bibinfo {author} {\bibfnamefont {C.}~\bibnamefont {Henkel}}, \bibinfo
  {author} {\bibfnamefont {K.}~\bibnamefont {Joulain}}, \ and\ \bibinfo
  {author} {\bibfnamefont {J.-J.}\ \bibnamefont {Greffet}},\ }\href@noop {}
  {\bibfield  {journal} {\bibinfo  {journal} {Physical Review B}\ }\textbf
  {\bibinfo {volume} {77}},\ \bibinfo {pages} {035431} (\bibinfo {year}
  {2008})}\BibitemShut {NoStop}%
\bibitem [{\citenamefont {Biehs}\ \emph {et~al.}(2007)\citenamefont {Biehs},
  \citenamefont {Reddig},\ and\ \citenamefont {Holthaus}}]{biehs2007thermal}%
  \BibitemOpen
  \bibfield  {author} {\bibinfo {author} {\bibfnamefont {S.-A.}\ \bibnamefont
  {Biehs}}, \bibinfo {author} {\bibfnamefont {D.}~\bibnamefont {Reddig}}, \
  and\ \bibinfo {author} {\bibfnamefont {M.}~\bibnamefont {Holthaus}},\
  }\href@noop {} {\bibfield  {journal} {\bibinfo  {journal} {The European
  Physical Journal B}\ }\textbf {\bibinfo {volume} {55}},\ \bibinfo {pages}
  {237} (\bibinfo {year} {2007})}\BibitemShut {NoStop}%
\bibitem [{\citenamefont {Basu}\ and\ \citenamefont
  {Wang}(2013)}]{basu2013near}%
  \BibitemOpen
  \bibfield  {author} {\bibinfo {author} {\bibfnamefont {S.}~\bibnamefont
  {Basu}}\ and\ \bibinfo {author} {\bibfnamefont {L.}~\bibnamefont {Wang}},\
  }\href@noop {} {\bibfield  {journal} {\bibinfo  {journal} {Applied Physics
  Letters}\ }\textbf {\bibinfo {volume} {102}},\ \bibinfo {pages} {053101}
  (\bibinfo {year} {2013})}\BibitemShut {NoStop}%
\bibitem [{\citenamefont {Moran}\ \emph {et~al.}(2009)\citenamefont {Moran},
  \citenamefont {Wainerdi}, \citenamefont {Cherukuri}, \citenamefont
  {Kittrell}, \citenamefont {Wiley}, \citenamefont {Nicholas}, \citenamefont
  {Curley}, \citenamefont {Kanzius},\ and\ \citenamefont
  {Cherukuri}}]{moran2009size}%
  \BibitemOpen
  \bibfield  {author} {\bibinfo {author} {\bibfnamefont {C.~H.}\ \bibnamefont
  {Moran}}, \bibinfo {author} {\bibfnamefont {S.~M.}\ \bibnamefont {Wainerdi}},
  \bibinfo {author} {\bibfnamefont {T.~K.}\ \bibnamefont {Cherukuri}}, \bibinfo
  {author} {\bibfnamefont {C.}~\bibnamefont {Kittrell}}, \bibinfo {author}
  {\bibfnamefont {B.~J.}\ \bibnamefont {Wiley}}, \bibinfo {author}
  {\bibfnamefont {N.~W.}\ \bibnamefont {Nicholas}}, \bibinfo {author}
  {\bibfnamefont {S.~A.}\ \bibnamefont {Curley}}, \bibinfo {author}
  {\bibfnamefont {J.~S.}\ \bibnamefont {Kanzius}}, \ and\ \bibinfo {author}
  {\bibfnamefont {P.}~\bibnamefont {Cherukuri}},\ }\href@noop {} {\bibfield
  {journal} {\bibinfo  {journal} {Nano Research}\ }\textbf {\bibinfo {volume}
  {2}},\ \bibinfo {pages} {400} (\bibinfo {year} {2009})}\BibitemShut {NoStop}%
\bibitem [{\citenamefont {Kheirandish}\ and\ \citenamefont
  {Ameri}(2014)}]{kheirandish2014electromagnetic}%
  \BibitemOpen
  \bibfield  {author} {\bibinfo {author} {\bibfnamefont {F.}~\bibnamefont
  {Kheirandish}}\ and\ \bibinfo {author} {\bibfnamefont {V.}~\bibnamefont
  {Ameri}},\ }\href@noop {} {\bibfield  {journal} {\bibinfo  {journal}
  {Physical Review A}\ }\textbf {\bibinfo {volume} {89}},\ \bibinfo {pages}
  {032124} (\bibinfo {year} {2014})}\BibitemShut {NoStop}%
\bibitem [{\citenamefont {Mitsui}\ and\ \citenamefont
  {Aoki}(2013)}]{mitsui2013observation}%
  \BibitemOpen
  \bibfield  {author} {\bibinfo {author} {\bibfnamefont {T.}~\bibnamefont
  {Mitsui}}\ and\ \bibinfo {author} {\bibfnamefont {K.}~\bibnamefont {Aoki}},\
  }\href@noop {} {\bibfield  {journal} {\bibinfo  {journal} {The European
  Physical Journal D}\ }\textbf {\bibinfo {volume} {67}},\ \bibinfo {pages}
  {213} (\bibinfo {year} {2013})}\BibitemShut {NoStop}%
\bibitem [{\citenamefont {Palik}(1998)}]{palik1998handbook}%
  \BibitemOpen
  \bibfield  {author} {\bibinfo {author} {\bibfnamefont {E.~D.}\ \bibnamefont
  {Palik}},\ }\href@noop {} {\emph {\bibinfo {title} {Handbook of optical
  constants of solids}}},\ Vol.~\bibinfo {volume} {3}\ (\bibinfo  {publisher}
  {Academic press},\ \bibinfo {year} {1998})\BibitemShut {NoStop}%
\bibitem [{\citenamefont {Manjavacas}\ and\ \citenamefont
  {de~Abajo}(2012)}]{manjavacas2012radiative}%
  \BibitemOpen
  \bibfield  {author} {\bibinfo {author} {\bibfnamefont {A.}~\bibnamefont
  {Manjavacas}}\ and\ \bibinfo {author} {\bibfnamefont {F.~J.~G.}\ \bibnamefont
  {de~Abajo}},\ }\href@noop {} {\bibfield  {journal} {\bibinfo  {journal}
  {Physical Review B}\ }\textbf {\bibinfo {volume} {86}},\ \bibinfo {pages}
  {075466} (\bibinfo {year} {2012})}\BibitemShut {NoStop}%
\end{thebibliography}%
\bibliographystyle{apsrev4-1}
\end{document}